\documentclass[12pt,english]{article}
\usepackage[T1]{fontenc}
\usepackage{amsmath}
\usepackage{amssymb}
\usepackage{graphicx}
\usepackage{subcaption}
\usepackage{booktabs} 
\usepackage{array}    
\usepackage{tabularx} 

\usepackage{graphicx}
\graphicspath{ {images/} }

\makeatletter
\usepackage{jheppub}
\def\@fpheader{\relax}
\allowdisplaybreaks[1]

\makeatother

\usepackage{babel}

\subheader{}

\title{Energy-time uncertainty relation from entropy measures}
\author[a]{Nana Cabo Bizet,}
\author[a]{Octavio Obreg\'on,}
\author[a]{Wilfredo Yupanqui Carpio.}

\affiliation[a]{Departamento de F\'isica, Divisi\'on de Ciencias e Ingenier\'ias, Universidad de Guanajuato, Loma del Bosque 103, Le\'on 37150, Guanajuato, M\'exico.}

\emailAdd{nana@fisica.ugto.mx}
\emailAdd{octavio@fisica.ugto.mx}
\emailAdd{w.yupanquicarpio@ugto.mx}

\abstract{In a previous study, it was shown that the Generalized Uncertainty Principle (GUP) can be derived from non-extensive entropies, particularly those depending only on the probability, denoted as $S_\pm$ in the literature. This finding reveals an intriguing connection between non-extensive statistics and quantum gravity. In the present work, we extend our previous result and derive a generalized energy-time uncertainty relation based on a measure of non-extensive entropies. Consequently, the dispersion relation undergoes modifications consistent with those obtained in other approaches to quantum gravity. We interpret these modifications as evidence of the non-extensive behavior of spacetime fluctuations at scales close to the Planck scale. While these effects are significant in this regime, they become negligible in the classical one, i.e. at low energies where the spacetime is smooth. As a consequence of the non-extensive behavior exhibited by spacetime at very small scales, the black hole radiation temperature undergoes quantum-level corrections, increasing in the case  of \( S_{-} \) and decreasing for the case of \( S_{+} \). Moreover, the modified uncertainty relation derived here predicts a maximum uncertainty in energy, of the order of Planck energy, and a minimum time interval, of the order of the Planck time, offering new insights into the fundamental structure of spacetime in the quantum regime.}

\begin{document}

\maketitle
\flushbottom

\section{Introduction}
\label{introduction}
Position $q$ and momentum $p$ are two complementary quantities of a quantum particle, satisfying an uncertainty relation first derived by H. Kennard \cite{kennard1927quantenmechanik} in 1927, and H. Weyl \cite{guth1929gruppentheorie} in 1929, mathematically represented by
\begin{equation}
    \Delta q \Delta p \geq \frac{\hbar}{2}, \label{eq:Uncert_relat_q_p_usual}
\end{equation}
where $\Delta q$ and $\Delta p$ represent the standard deviation of the position and momentum, respectively, and $\hbar$ is the reduced Planck constant. This equation expresses Heisenberg's Uncertainty Principle (HUP), which states that it is not possible to simultaneously determine the position and momentum of a quantum particle. The same expression is obtained by considering position and momentum as operators that satisfy the commutation relation
\begin{equation}
    [\hat{q}, \hat{p}] = i\hbar. \label{eq:Conmut_rel_x_p}
\end{equation}
Substituting this into Robertson's uncertainty relation (1929) \cite{PhysRev.34.163},
\begin{equation}
    \Delta A \Delta B \geq \frac{1}{2} \left| \langle \psi | [\hat{A}, \hat{B}] | \psi \rangle \right|, \label{eq:General_Heisenbrg_ineqlty}
\end{equation}
we obtain \eqref{eq:Uncert_relat_q_p_usual}, where $|\psi\rangle$ is the physical state of the system. . Additionally, \( A \) and \( B \) are operators corresponding to observables.

Heisenberg also postulated the energy-time uncertainty relation, expressed mathematically as
\begin{equation}
    \Delta t \Delta E \geq \frac{\hbar}{2},\label{enq.time_energy_uncertainty_relation}
\end{equation}
where $\Delta E$ represents the energy dispersion and $\Delta t$ is defined as a specific time interval. The inequality \eqref{enq.time_energy_uncertainty_relation} cannot be interpreted in the same way as the well-known Heisenberg relation between position and momentum \eqref{eq:Uncert_relat_q_p_usual}. This distinction arises because in quantum mechanics, time is regarded as an evolution parameter rather than as an observable. Hence, it cannot be represented by a self-adjoint operator (conjugated to the Hamiltonian) as argued by Pauli due to the bounded nature of the continuous energy spectrum \cite{pauli2012general}. The time-energy uncertainty relation was initially deduced in the work by Mandelstam and Tamm \cite{Mandelstam1991}, and subsequently discussed by various authors \cite{WANG20072304, 10.1119/1.18410, 10.1119/1.18880, KIJOWSKI1974361, olkhovsky1974time, 10.1119/1.1973651, 10.1143/PTP.66.1525, Dodonov_2015, 10.1119/1.1430697}.

There exist several methods to derive the energy-time uncertainty relation, \eqref{enq.time_energy_uncertainty_relation}. Particularly noteworthy is the approach presented in \cite{PhysRevA.50.933}, which provides a derivation from first principles involving a canonical transformation in classical mechanics. This transformation yields a new canonical momentum (the energy $E$) and a new canonical coordinate $T$ (referred to as tempus), which is conjugate to the energy. It is crucial to emphasize that tempus, as a new canonical coordinate, is conceptually distinct from the time $t$ that denotes the system's evolution. The canonical invariance of the Poisson bracket ensures that energy and tempus follow the same algebraic relationships as momentum and position. Upon quantization of the system, the uncertainty relation \eqref{enq.time_energy_uncertainty_relation} emerges.

Introducing the quantum effects of gravity, the uncertainty relation for position and momentum, \eqref{eq:Uncert_relat_q_p_usual}, is extended to a Generalized Uncertainty Principle (GUP), which ensures a nonzero minimal uncertainty in position \cite{G_Veneziano_1986,scardigli1995some,BAGCHI20094307}. In \cite{BIZET2023137636}, it is demonstrated that the GUP arises from the consideration of non-extensive entropies. In particular they arise for the entropis denoted as $S_\pm$. The deformation parameters, which determine the magnitude of minimal uncertainty effects, for the measures $S_{+}$ and $S_{-}$, are negative and positive, respectively. This enables exploration of various possibilities in the quest for physical implications. Therefore, non-extensive statistics can be interpreted as a signature of quantum gravity.

In recent years some works have explored the relation between quantum gravity and non-extensivity as well from different perspectives \cite{PhysRevD.103.026021,MAJHI201732,PhysRevD.105.044042}.

GUP is also derived from different proposals: in \cite{G_Veneziano_1986}, the scattering of strings at ultra-high energies is considered to analyze the divergences of quantum gravity at the Planck scale; in \cite{MAGGIORE199365}, a gedanken experiment is proposed to measure the area of the apparent horizon of black holes in the context of quantum gravity; and \cite{SCARDIGLI199939} explores the idea that spacetime in the Planck region fluctuates, leading to the possibility of virtual micro-black holes affecting the measurement process.

The non-extensive entropies $S_\pm$, first introduced in \cite{e12092067}, were derived within the framework of Superstatistics \cite{PhysRevE.88.062146, doi:10.1142/S0217751X15300392}. These entropies depend only on the probabilities associated with the microscopic configuration of the system and enable exploration of non-equilibrium phenomena. These generalized entropies represent the entire family of entropies dependent solely on probability \cite{PhysRevE.88.062146}. The statistics associated with $S_{+}$ and $S_{-}$ exhibit distinct characteristics: the former gives rise to an effective potential related to an effective repulsive contribution term, while the latter contributes effectively attractively \cite{GILVILLEGAS2017364}.

In this work, we expand upon the results obtained in \cite{BIZET2023137636}. As observed, the generalized entropies ensuring quantum gravity effects lead to modifications in the uncertainty relation and, consequently, in the algebra between position and momentum. Similarly, these entropies are expected to modify the uncertainty relation between time and energy, \eqref{enq.time_energy_uncertainty_relation}. 

The structure of the paper is as follows: In section \ref{sec:Derivation_of_the_energy_time_uncertainty_relation}, we present the derivation of the uncertainty relation between energy and time by introducing a canonical variable called tempus, which is conjugate to energy. In section \ref{sec:Modified_entropies_and_the_Generalized_Uncertainty_Principle}, we discuss the origin of non-extensive entropies $S_\pm$ in the domain of Superstatistics and how GUP emerges from them. In section \ref{sec:Generalized_energy_time_uncertainty_relation}, we first derive the the  energy-time uncertainty relation from Scardigli's arguments. Then, we derive the same uncertainty from non-extensive entropies, by modifying the commutation relation between the tempus operator and energy. We present our conclusions in section \ref{sec:Conclusions}.

\section{Derivation of the energy-time uncertainty relation}
\label{sec:Derivation_of_the_energy_time_uncertainty_relation}
A rigorous derivation of the uncertainty relation between time and energy was first provided by \cite{Mandelstam1991}. In this work, we will follow a different approach, as presented by \cite{PhysRevA.50.933}, where a new canonical variable, called tempus $T$, conjugate to energy $E$ is defined through a canonical transformation. In this part of the article, we will briefly review this approach.

The Hamiltonian  $H=H(q,p,t)$, is a function that describes a system in terms of its position $q$, momentum $p$, and time $t$. It describes the time development of the system through Hamilton's equations
\begin{align}
    \dot{q}&=\frac{\partial H}{\partial p}, &\dot{p}&=-\frac{\partial H}{\partial q}, \label{eq:Hamlt_equas}
\end{align}
where $\dot{q}$ and $\dot{p}$ represent the derivative of position and momentum with respect to time, respectively. It is important to note that the energy of a time-dependent system does not necessarily equal to its Hamiltonian \cite{DHKobe_1987}. The energy \( E(q, \dot{q}, t) \) is a function of the generalized coordinate \( q \), the generalized velocity \( \dot{q} \), and time \( t \). It is generally expressed as the sum of the kinetic energy and the conservative potential energy. In contrast, the Hamiltonian is derived from the Lagrangian through the canonical procedure. Moreover, the Hamiltonian governs the system's dynamics via Hamilton's equations \eqref{eq:Hamlt_equas}. However, it is not gauge-invariant and therefore cannot generally be measured \cite{10.1119/1.17334}. On the other hand, in principle, energy differences can be measured. Consequently, it is essential to distinguish between the energy and the Hamiltonian of the system \cite{DHKobe_1987}.

By a canonical transformation one could find a coordinate transformation from the set ($q$, $p$) to a new set of canonical variables ($Q$, $P$), which also satisfies Hamilton's equations. We can choose the energy \( E \) as the new canonical momentum \( P \), while the new generalized coordinate \( Q \), conjugate to the energy, is denoted by the canonical variable called \emph{tempus} \( T \) \cite{DHKobe_1987}. This new variable, conjugate to the energy \( E \), generally differs from the time \( t \), which governs the system's evolution \cite{PhysRevA.50.933}. Since the transformation is canonical, the old canonical momentum can be expressed in terms of the generating function of the second type, \( S(q, E, t) \), as
\begin{equation}
    p=\frac{\partial S(q,E,t)}{\partial q},\label{eq:p_partial_diff_S}
\end{equation}
and the canonical variable \emph{tempus} can be obtained from
\begin{equation}
    T=\frac{\partial S(q,E,t)}{\partial E}.\label{eq:T_partial_diff_S}
\end{equation}
By integration $p$ in \eqref{eq:p_partial_diff_S}, the generating function $S$ can be obtained
\begin{eqnarray}
    S\left(q,E,t\right)=\int_{q_0}^{q}p\left(q',E,t\right)dq'+S\left(q_0,E,t\right),\label{eq.S_integrate_from_p}
\end{eqnarray}
where $q_0$ is an arbitrary initial displacement. To solve this integral, we need to determine the old canonical momentum \( p \) as a function of \( q \), \( E \), and \( t \). The arbitrary function \( S(q_0, E, t) \) does not affect the system's dynamics and can often be set to zero without loss of generality \cite{10.1119/1.17334}. On the other hand, the new generalized coordinate, \emph{tempus} \( T \), can be expressed as a function of \( q \), \( E \) and \( t \) applying the partial differentiation of equation \eqref{eq.S_integrate_from_p} with respect to the energy, as derived from equation \eqref{eq:T_partial_diff_S}.

As previously mentioned, \( T \) and \( E \) are canonical variables that satisfy Hamilton's equations
\begin{align}
    \dot{T}&=\frac{\partial H'}{\partial E}, &\dot{E}&=-\frac{\partial H'}{\partial T}.\label{eq:Hamilton_eq_for_T_E}
\end{align}
The new Hamiltonian \( H' \) is expressed in terms of the old Hamiltonian \( H \) and the generating function \( S \) as \( H' = H + (\partial S / \partial t) \). Since the energy \( E = E(q, \dot{q}, t) \) and the Hamiltonian \( H = H(q, p, t) \) are not necessarily equal, the function \( \Phi(q, p, t) = H - E \) is defined to represent their difference \cite{10.1119/1.17334}. When this difference is included in the new Hamiltonian \( H' \), Hamilton's equations \eqref{eq:Hamilton_eq_for_T_E} take the form
\begin{align}
    \dot{T}&=1+\frac{\partial}{\partial E}\left(\Phi+\frac{\partial S}{\partial t}\right), \label{eq:T_EQUAT_OF_MOTION_Phi_S} \\ 
    \dot{E}&=-\frac{\partial}{\partial T}\left(\Phi+\frac{\partial S}{\partial t}\right). \label{eq:E_EQUAT_OF_MOTION_Phi_S}
\end{align}
In both equations, it is necessary to express \( \Phi + (\partial S / \partial t) \) as a function of \( E \), \( T \), and \( t \) before performing the differentiation. The equations \eqref{eq:T_EQUAT_OF_MOTION_Phi_S} and \eqref{eq:E_EQUAT_OF_MOTION_Phi_S} can then be solved for \( T(t) \) and \( E(t) \), respectively, with the solution to the original problem given by \( q = q(E(t), T(t), t) = q(t) \). In addition, these equations are invariant under gauge transformations \cite{10.1119/1.17334}.

For a conservative system, it is possible to choose a gauge such that the generating function \( S \) does not explicitly depend on time, i.e., \( \partial S / \partial t = 0 \), and the Hamiltonian becomes equal to the energy, \( H = E \) \cite{10.1119/1.17334}. This implies that \( \Phi = 0 \), which consequently reduces Hamilton's equations \eqref{eq:T_EQUAT_OF_MOTION_Phi_S} and \eqref{eq:E_EQUAT_OF_MOTION_Phi_S} to trivial solutions
\begin{align}
T&=t-t_0, &E&=E_0, \label{eq:Sol_eq_for_T_E_Conserv}
\end{align}
where $t_0$ and $E_0$ are arbitrary constants. The solution for \( T \) in equation \eqref{eq:Sol_eq_for_T_E_Conserv} shows that the canonical variable \emph{tempus}, in this particular case, simplifies to the time difference \( t - t_0 \). The solution to the original problem is therefore $q=q(E_0,t-t_0)=q(t)$.

\subsection{Quantization}
The transition from classical mechanics to quantum mechanics is achieved by promoting the dynamic variables to quantum operators and replacing the Poisson bracket with a commutator, e.g. $\{\ ,\ \}\rightarrow\frac{1}{i\hbar}[\ ,\ ]$. For the position and momentum operators, the commutation relation they satisfy is given in equation \eqref{eq:Conmut_rel_x_p}, and the corresponding uncertainty relation is expressed in equation \eqref{eq:Uncert_relat_q_p_usual}. Similarly, one can define the \emph{tempus} operator \( \hat{T} \) and the energy operator \( \hat{E} \) (because these variables are canonically conjugate), which satisfy the commutation relation
\begin{eqnarray}
    [\hat{T},\hat{E}]=i\hbar.\label{eq.Usuall_Commut_rela_T_E}
\end{eqnarray}
From this, the corresponding uncertainty relation for the \emph{tempus} and energy operators can be derived using the Robertson inequality \eqref{eq:General_Heisenbrg_ineqlty}
\begin{eqnarray}
   \Delta T\Delta E\geq\frac{\hbar}{2}.\label{enq.Tempus_energy_uncertainty}
\end{eqnarray}
This uncertainty relation is defined on the same solid foundation as the momentum-position uncertainty relation \eqref{eq:Uncert_relat_q_p_usual}. This is made possible by choosing the energy \( E \) as a new canonical momentum and defining a new canonical coordinate \( T \), which, unlike \( t \), is the conjugate operator to the energy operator \cite{PhysRevA.50.933}.

A measure of how fast the system is evolving can be obtained by computing the time derivative of the expected value of an observable \( Q \) using Ehrenfest's theorem
\begin{equation}
    \frac{d\langle Q\rangle}{dt}=\frac{1}{i\hbar}\langle[\hat{Q},\hat{H}]\rangle+\langle\frac{\partial \hat{Q}}{\partial t}\rangle.\label{eq:Erenfest_theorem_Q}
\end{equation}
If the operator \( \hat{Q} \), representing an observable, does not explicitly depend on time \( t \), the rate of change of its expected value is determined solely by its commutator with the Hamiltonian. By choosing \( A = H \) and \( B = Q \) in the generalized uncertainty relation \eqref{eq:General_Heisenbrg_ineqlty}, and assuming that \( Q \) does not explicitly depend on \( t \), we obtain \cite{griffiths2019introduction}
\begin{eqnarray}
   \Delta Q\Delta H\geq\frac{\hbar}{2}\left|\frac{d\langle Q\rangle}{dt}\right|.\label{enq.Uncertainty_Q_H}
\end{eqnarray}
As before, for a conservative system, one can choose \( \hat{E} = \hat{H} \), where the Hamiltonian operator is time-independent \cite{PhysRevA.50.933}. The time dependence, in the chosen picture, lies within the wavefunction that describes the system. We also define \cite{griffiths2019introduction}
\begin{eqnarray}
    \Delta t\equiv\frac{\Delta Q}{|d\langle Q\rangle/dt|}.\label{eq:Delta_t_Delta_T_velocty}
\end{eqnarray}
This equation establishes that the uncertainty in \( Q \) is related to the rate at which its expected value changes over a given time interval \( \Delta t \). Qualitatively, if the expected value of an observable varies rapidly over time, its uncertainty will be greater for the same time interval.

Setting \( T = Q \) in \eqref{eq:Delta_t_Delta_T_velocty} yields \( \Delta t = \Delta T \), according to \eqref{eq:Erenfest_theorem_Q} for the case where \( \hat{E} = \hat{H} \) and the \emph{tempus} operator does not explicitly depend on time. Substituting this into \eqref{enq.Tempus_energy_uncertainty} or \eqref{enq.Uncertainty_Q_H}, we arrive at the energy-time uncertainty relation \eqref{enq.time_energy_uncertainty_relation}. The procedure followed in this section provides a rigorous derivation of the energy-time uncertainty relation based on first principles \cite{PhysRevA.50.933}. Reference \cite{PhysRevA.50.933} shows that the expectation value of the tempus operator is directly related to the time $t$ of the system's evolution, which directly leads, from \eqref{eq:Delta_t_Delta_T_velocty}, to \eqref{enq.time_energy_uncertainty_relation}.

Another alternative way to obtain \eqref{enq.time_energy_uncertainty_relation} is from the uncertainty relationship for position and momentum \eqref{eq:Uncert_relat_q_p_usual}, considering $\Delta E=v_{G}\Delta p$ and $\Delta q=v_{G}\Delta t$, where $v_{G}$ is the group velocity of the wave packet. Considering these factors, an alternative derivation of the uncertainty relation between energy and time is obtained.

The \emph{tempus} operator can be constructed from its corresponding classical expression \eqref{eq:T_partial_diff_S}. For example, in \cite{PhysRevA.50.933}, such an operator is derived for both a free particle and a particle under the influence of a constant force. To construct a self-adjoint \emph{tempus} operator, it must be expressed in terms of the position \( \hat{q} \) and the momentum \( \hat{p} \) operators \cite{PhysRevA.50.933}. Consequently, the \emph{tempus} operator \( \hat{T} \) can be represented either in position space or momentum space, depending on the physical problem. For further details, the reader is referred to \cite{PhysRevA.50.933}.

\section{Modified entropies and the Generalized Uncertainty Principle}
\label{sec:Modified_entropies_and_the_Generalized_Uncertainty_Principle}
In the Heisenberg uncertainty relation, the gravitational interaction between particles was completely neglected, although this was somehow justified by the considerable weakness of gravity when compared with other fundamental interactions.
Incorporating gravity into quantum measurement processes led to the generalization of the uncertainty relation \eqref{eq:Uncert_relat_q_p_usual}. This modification is known as the Generalized Uncertainty Principle (GUP), which introduces additional terms that account for gravitational effects at very small scales, typically at the Planck scale.\par

\subsection{GUP from modified entropies}
GUP can alternatively be derived from modified entropies \cite{BIZET2023137636}, in particular of the ones that depend only on the probability \cite{e12092067}. These entropies are obtained in the Superstatistics formalism \cite{doi:10.1142/S0217751X15300392}, mathematically expressed as
\begin{eqnarray}
    S_{+}&=&k_{\beta}\sum_{l}^{\Omega}\left(1-p_l^{p_l}\right),\label{Eqq.Entropy.S_+}\\
    S_{-}&=&k_{\beta}\sum_{l}^{\Omega}\left(p_l^{-p_l}-1\right),\label{Eqq.Entropy.S_-}
\end{eqnarray}
and in terms of the Boltzmann-Gibbs (BG) (Shannon) entropy $S_{BG}=k_{\beta}\ln{\Omega}$, for $p_l=\frac{1}{\Omega}$, can be written as (considering the expansion)
\begin{equation}
    S_{\pm}=S_{BG}\mp\frac{1}{2!}S_{BG}^2 e^{-S_{BG}}+\cdots.\label{S+-.Entro.SB}
\end{equation}
The associated probabilities and, consequently, the exponential and logarithms are also generalizations of the usual ones; the exponentials result in
   \begin{equation}
       \exp_\pm(-x)=\exp(-x)\sum_{j=0}^{\infty}a^\pm_jx^j,\label{GeneralizedExponential}
   \end{equation}
   where the dimensionless coefficients ${a}^\pm_j$ are fixed \cite{BIZET2023137636}. One can then define in space of phase a generalized probability
   \begin{equation}
       \mathcal{P}_\pm=\exp(-\bar{H})\sum_{j=0}^{\infty}\bar{a}^\pm_j\bar{H}^j,\label{6}
   \end{equation}
being $\bar H$ the dimensionless Hamiltonian.  This is a series expansion that has a convergence radius of order 1, that we denote $R_{\pm}=\lim_{n\rightarrow \infty} \frac{\bar a^{\pm}_n}{\bar a^{\pm}_{n+1}}$. One can invert (\ref{6}) to obtain $H_{\pm}$
in terms of $\bar{H}$, then the convergence radius is further restricted by the logarithm expansion. One gets the generalized Hamiltonian expansions \cite{BIZET2023137636}
\begin{equation}
    H_\pm=\frac{1}{2m}\left(p^2+\frac{\alpha_\pm}{3}p^4+\cdots\right),\label{eq:Eff_Hamiltonian}
\end{equation}
in terms of low-energy momentum $p$. Then, it is possible to define a generalized momenta $p_\pm$ in terms of $p$, as follows
\begin{eqnarray}
p_\pm=p+\frac{\alpha_\pm}{3} p^3+\cdots,\label{Eqq.Eff.Moment.S_{+-}}
\end{eqnarray}
where, $\alpha_\pm=\alpha^0_\pm/M_{x}^2$ ($c=1$ is assumed) is the deformed parameter, $M_{x}$ is a certain energy scale, and $\alpha^0_\pm$ is a dimensionless constant
\begin{eqnarray}
\alpha_\pm^{0}=\frac{3 \left(\left(a^\pm_1\right)^2-2{a}^\pm_2\right)}{8 \left(1-{a}^\pm_1\right)}.\label{Eq.Dimenssles.Deford.Paramter}
\end{eqnarray}
The commutation relation between $\hat{p}_\pm$ and $\hat{q}$ can be calculated using equation \eqref{Eqq.Eff.Moment.S_{+-}}, and the usual commutation relation $\left[\hat{q},\hat{p}\right]=i\hbar$
\begin{eqnarray}
\left[\hat{ q},\hat{p}_\pm\right]=i\hbar \left(1+\alpha_\pm \hat{p}^2+\cdots\right),\label{Modif.Conmut.Relat.Diment.x.p_pm}
\end{eqnarray}
taking into account that $\alpha_\pm \hat{p}_{\pm}^2=\alpha_\pm \hat{p}^2+\cdots$, it allows us to express \eqref{Modif.Conmut.Relat.Diment.x.p_pm} as
\begin{eqnarray}
\left[\hat{ q},\hat{p}_\pm\right]=i\hbar \left(1+\alpha_\pm\ \hat p_\pm^2\right).\label{Modif.Conmut.Relat.Diment.x.p_pm.Fam}
\end{eqnarray}
A compact way to express the modified momentum $p_\pm$ in \eqref{Eqq.Eff.Moment.S_{+-}}, which satisfies the commutation relation \eqref{Modif.Conmut.Relat.Diment.x.p_pm.Fam}, is
\begin{equation}
    p_\pm=\frac{1}{\sqrt{\alpha_\pm}}\tan{(\sqrt{\alpha_\pm}p)},\label{eq:Compact_p_effect}
\end{equation}
where the domain of $p$ is restricted to $-\pi/2\sqrt{\alpha_\pm}<p<\pi/2\sqrt{\alpha_\pm}$. It is essential to mention that the coordinate $q$ and the effective momentum $p_\pm$ are no longer canonical conjugates \cite{PhysRevD.52.1108}.\par

The generalized uncertainty relation associated with the modified algebra in \eqref{Modif.Conmut.Relat.Diment.x.p_pm.Fam} is
\begin{eqnarray}
    \Delta q\Delta p_\pm\geq\frac{\hbar}{2}\left(1+\alpha_\pm\left(\Delta p_\pm\right)^2\right),\label{enq.GUP_q_p}
\end{eqnarray}
which is a modification of the uncertainty relation in \eqref{eq:Uncert_relat_q_p_usual}, taking into account quantum corrections of gravity. The parameter $\alpha_\pm^{0}$ can have positive and negative values, depending on the considered statistics, i.e.,  $\alpha^0_+=-0.560565$ and $\alpha^0_{-}=0.361022$ \cite{BIZET2023137636}.  When the positive value is set, \eqref{enq.GUP_q_p} introduce a minimal length in the position, $(\Delta q)_{min}=\hbar\sqrt{\alpha_-}$. If we choose the Planck scale $M_x=M_{Pl}$ the parameter $\alpha^0_-$ is in order of unity and the minimal uncertainty in the position becomes the Planck length $l_{Pl}$.  On the other hand, when the negative value is chosen, a maximum uncertainty in the momentum is obtained, $(\Delta p)_{\text{max}}\sim 1/\sqrt{|\alpha_{+}|}$.

\subsection{GUP from a micro-black hole gedanken experiment}
As mentioned in the introduction of this work, GUP can be derived from different arguments. In \cite{SCARDIGLI199939}, GUP is derived by considering the measurement process, under the effects of gravity. At the Planck scale, spacetime is believed to undergo significant metric fluctuations, allowing for the possible formation of numerous virtual micro black holes. We briefly explain this process.

According to Heisenberg's uncertainty principle, in the high-energy approximation $\Delta E\Delta q\geq \hbar/2$, when attempting to observe a region of space with width $\Delta q$, the spacetime metric in that region is expected to undergo quantum fluctuations with an energy amplitude given by $\Delta E \sim \hbar/2\Delta q$, confined within the region of width $\Delta q$. The Schwarzschild radius $R=2G \Delta E$, associated with the energy amplitude $\Delta E$, generally falls within the region $\Delta q$ for small fluctuations in energy. As the observation region $\Delta q$ decreases, the energy fluctuation $\Delta E$ increases, and consequently, $R$ grows until it eventually matches the width of $\Delta q$, leading to the formation of a micro black hole; this occurs when $\Delta q$ searches the Planck length \cite{scardigli1995some}. To observe finer details, an energy greater than the Planck energy $M_{Pl}$ must be concentrated in that region, which would further increase the Schwarzschild radius $R$, obscuring more details of the observed region due to the expansion of the micro black hole's event horizon.

Since $R$ depends linearly on $\Delta E$, while $\Delta q$ depends inversely, we can combine both conditions into a single expression \cite{SCARDIGLI199939}
\begin{equation}
    \Delta q\geq\frac{\hbar}{2\Delta E}+2G\Delta E.\label{enq:Delta_x_vs_Delta_E_Scardigli}
\end{equation}
If the energy fluctuation $\Delta E$ reaches the Planck energy $M_{Pl}$ in \eqref{enq:Delta_x_vs_Delta_E_Scardigli}, a minimum observable region with a width on the order of the Planck length $L_{Pl}$ emerges. This can be demonstrated by considering that $G M_{Pl} = L_{Pl}$ and $\hbar G = L_{Pl}$ \cite{scardigli1995some}.

In the high-energy regime, we have $\Delta E \sim \Delta p$, which in \eqref{enq:Delta_x_vs_Delta_E_Scardigli} can be expressed as
\begin{equation}
    \Delta q\Delta p\geq\frac{\hbar}{2}+\frac{2L_{Pl}^2}{\hbar}(\Delta p)^2,\label{enq:Dq_Dp_GUP_Scardigli}
\end{equation}
and represents a generalization of the uncertainty principle to cases where gravity is significant and energies are on the order of the Planck scale. This result is similar to the one obtained from entropic principles in \eqref{enq.GUP_q_p} for the positive deformation parameter $\alpha_{-}$. This suggests a connection between nonextensive statistics and quantum gravity, implying that quantum gravity exhibits inherently nonextensive behavior. It reinforces our argument that non-extensivity effects are significant at scales near the Planck regimen. Notice that the relation (\ref{enq:Dq_Dp_GUP_Scardigli}) can be solved
to find inferior and superior bounds for $\Delta p$ as has been shown in \cite{PhysRevD.52.1108}.

The difference between the results obtained in \eqref{enq:Dq_Dp_GUP_Scardigli} and \eqref{enq.GUP_q_p}, aside from the different principles from which they are derived, is that the statistics related to entropy $S_{+}$ lead to a maximum uncertainty in momentum, a feature not observed in \eqref{enq:Dq_Dp_GUP_Scardigli}.

\section{Generalized energy-time uncertainty relation}
\label{sec:Generalized_energy_time_uncertainty_relation}
It has been demonstrated that starting from non-extensive entropies, it is possible to obtain a generalization to the uncertainty relation \eqref{eq:Uncert_relat_q_p_usual} to obtain \eqref{enq.GUP_q_p}. Using the same arguments, we can find a modified uncertainty relation between time and energy, which should be a generalization of \eqref{enq.time_energy_uncertainty_relation} to include quantum gravity corrections. Before proceeding, we will demonstrate that the modified energy-time uncertainty relation can be derived from Scardigli's arguments \cite{SCARDIGLI199939}.

\subsection{Generalized energy-time uncertainty relation from a micro-black hole gedanken experiment}
As mentioned in the previous section, it is impossible to observe the interior of the region occupied by a black hole due to the presence of the event horizon, a condition that persists throughout the black hole's entire lifetime. If gravitational effects are neglected, the energy-time uncertainty relation, $\Delta t \sim \hbar / 2\Delta E$, implies that the energy concentrated within a region of width $\Delta q$ will take a time $\Delta t$ to escape that region. As the energy fluctuation increases, the escape time decreases. However, when gravity is considered in the measurement process, a micro black hole forms, and consequently, an event horizon, as previously discussed. This event horizon prevents the energy from escaping the black hole; only a small amount escapes due to Hawking evaporation. As a result, the region inside the black hole remains unobservable for much longer than in the usual case. The duration for which the region inside the micro black hole remains unobservable is, evidently, equal to the lifetime of the micro black hole \cite{SCARDIGLI199939}. 

In general, for an arbitrary width $\Delta q$, the constraint on the time $\Delta t$ required for the energy $\Delta E$ to escape the region $\Delta q$ can be derived from relation \eqref{enq:Delta_x_vs_Delta_E_Scardigli} by considering $\Delta t \approx  \Delta q/v$ (in general $v\leq1$ and $v=c=1$ for the speed of light), yielding
\begin{equation}
    \Delta t\geq\frac{\hbar}{2\Delta E}+2G\Delta E.\label{enq:Delta_t_vs_Delta_E_new}
\end{equation}
Notice that  we have divided (\ref{enq:Delta_x_vs_Delta_E_Scardigli}) by $v$, and even for $v< 1$ the inequality (\ref{enq:Delta_t_vs_Delta_E_new}) is satisfied.
From which the generalized energy-time uncertainty relation
\begin{equation}
    \Delta t\Delta E\geq\frac{\hbar}{2}+2G(\Delta E)^2,\label{enq:Delta_t_Delta_E_GUP}
\end{equation}
is obtained. This expression accounts for gravity in measuring the lifetime of a state with energy amplitude $\Delta E$. From equation \eqref{enq:Delta_t_vs_Delta_E_new}, we observe that if the energy of the state is small, the first term on the right-hand side becomes more significant than the second, recovering the usual uncertainty relation. Conversely, as the energy increases, the first term becomes negligible compared to the second, and the lifetime of the state grows proportionally to its energy. If the energy fluctuation $\Delta E$ is the Planck energy, then from \eqref{enq:Delta_t_vs_Delta_E_new}, it can be observed that a minimum time interval on the order of the Planck time $\tau_{Pl}$ emerges, i.e., $(\Delta t)_{\text{min}} \sim \tau_{Pl}$. This is because the lifetime of a virtual black hole is approximately equal to the Planck time.

\subsection{Generalized energy-time uncertainty relation derived from modified entropies}
Before finding the uncertainty relation between time and energy from entropic measures, we must express the relativistic version of the effective Hamiltonian $H_{\pm}$ in \eqref{eq:Eff_Hamiltonian}. In the relativistic regime, energy $E$ and momentum $p$ are written as the components of a four-momentum vector $p_0=(E,p)$. Similarly, we should be able to write a modified four-momentum with components $P_\pm=(E_\pm,p_\pm)$, where $p_\pm$ is defined in \eqref{Eqq.Eff.Moment.S_{+-}}. On the other hand, $E_\pm$ must be the relativistic energy for the Hamiltonian $H_\pm$ in \eqref{eq:Eff_Hamiltonian}. We define this as
\begin{equation}
    E_\pm=E+\frac{\alpha_\pm}{3} E^3+\cdots.\label{Eqq.Eff.Energy.S_{+-}}
\end{equation}
It can be easily verified that this expression reduces to the Hamiltonian \eqref{eq:Eff_Hamiltonian} in the non-relativistic limit. With this purpose, we rewrite \eqref{Eqq.Eff.Energy.S_{+-}} as follows
\begin{equation}
    E_\pm=E\left(1+\frac{\alpha_\pm}{3} E^2+\cdots\right),\label{Eqq.Eff.Energy_reduc}
\end{equation}
and considering the relativistic energy relation express as $E= m$$\sqrt{1+p^2/m^2}$ in non-relativistic limit, we have
\begin{equation}
   E_\pm=m\left(1+\frac{p^2}{2m^2}\right)\left[1+\frac{\alpha_\pm}{3}p^2+\frac{\alpha_\pm}{3}m^2+\cdots\right], \label{Eint}
\end{equation}
which reduces to
\begin{equation}
    E_\pm=m\left(1+\frac{\alpha_\pm}{3}m^2\right)+\left(1+\alpha_\pm m^2\right)\frac{p^2}{2m}+\frac{\alpha_\pm}{3}\frac{p^4}{2m}+\cdots.\label{Eqq.Eff.Energy_reduc_3}
\end{equation}
As we have seen already, the deformation parameters $\alpha_\pm$ are expressed in terms of the Planck energy $M_{Pl}$ through the relation $\alpha_\pm=\alpha^0_\pm/M_{Pl}^2$. This allows us to rewrite \eqref{Eqq.Eff.Energy_reduc_3} as
\begin{multline}
      E_\pm=m\left(1+\frac{\alpha^0_\pm}{3}\left(\frac{m}{M_{Pl}}\right)^2\right)+\left(1+\alpha^0_\pm \left(\frac{m}{M_{Pl}}\right)^2\right)\frac{p^2}{2m}+\frac{\alpha_\pm}{3}\frac{p^4}{2m}+\cdots,\label{Eqq.Eff.Energy_reduc_4}
\end{multline}
where $\alpha^0_\pm$ is a dimensionless parameter defined in the previous section. On the other hand, $m$ represents the energy at low scales, implying that $M_{Pl}\gg m$, and therefore $m/M_{Pl}\ll 1$.  Notice that the conditions $m \ll M_{pl}$ have to be stronger than the condition $p \ll m$, i.e. $m/M_{Pl}\ll p/m$. This means that the mass of the particle has to be below the Planck scale, and therefore it can not be an excited state of string theory (for example). Considering this fact, we can write the effective energy \eqref{Eqq.Eff.Energy_reduc_4} in a familiar form
\begin{equation}
    E_\pm\simeq m+\frac{1}{2m}\left(p^2+\frac{\alpha_\pm}{3}p^4+\cdots\right).\label{Eqq.Eff.Energy_reduc_5}
\end{equation}
The first term represents the rest energy of the particle, and the second one represents the effective Hamiltonian \eqref{eq:Eff_Hamiltonian}. With this, we demonstrate that \eqref{Eqq.Eff.Energy.S_{+-}} is the relativistic expression of \eqref{eq:Eff_Hamiltonian}. The same expression can be obtained from the effective momenta \eqref{Eqq.Eff.Moment.S_{+-}} when a relatively high energy is applied to a particle with relatively low rest mass, it can be approximated as $E\simeq p$ in high-energy physics. Let us comment here on the convergence of the series (\ref{Eint}), (\ref{Eqq.Eff.Energy_reduc_3}) and (\ref{Eqq.Eff.Energy_reduc_5}). They have the same convergence radius of $H_{\pm}$, with the addition of the non-relativistic  condition $p^2\ll m^2$.

So, from \eqref{Eqq.Eff.Energy.S_{+-}} we have
\begin{equation}
    E_\pm=p+\frac{\alpha_\pm}{3} p^3+\cdots.\label{eq:MDR_entropy}
\end{equation}
This expression is interpreted as a modified dispersion relation (MDR) derived from entropic principles.

\begin{figure}[htb!] 
\begin{center}
\includegraphics[scale=0.8]{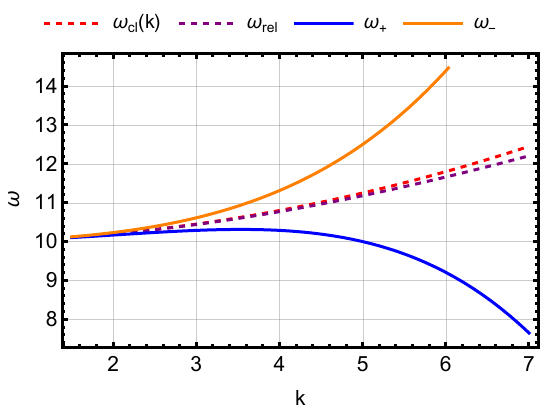} 
\caption{Graphical representation of equation \eqref{eq:Mod_Disp_Relt_Entrop}. In this graph, we compare the De Broglie dispersion relation, \( \omega_{cl} \), and the relativistic dispersion relation, \( \omega_{rel} \), with the modified dispersion relations \( \omega_{\pm} \).}
\label{fig:Plot_w_vs_k_compar}
\end{center}
\end{figure}

Building on other proposals in quantum gravity, similar MDRs to \eqref{eq:MDR_entropy} have been derived; for example, we cite \cite{nozari2006generalized,XIANG2006519,PhysRevLett.84.2318,Amelino-Camelia_2006,PhysRevD.70.107501}. In the non-relativistic approximation, the MDR in \eqref{eq:MDR_entropy} reduces, using \eqref{Eqq.Eff.Energy_reduc_5}, to
\begin{equation}
    \omega_{\pm}(k)=\frac{m}{\hbar}+\frac{\hbar k^2}{2m}\left(1+\alpha_{\pm}\frac{\hbar^2k^2}{2m}\right)+\cdots,\label{eq:Mod_Disp_Relt_Entrop}
\end{equation}
where the de Broglie relations for energy and momentum for matter waves $E = \hbar\omega$ and $p = \hbar k$, respectively, have been utilized. Clearly, this expression reduces to the usual dispersion relation $\omega_{cl}(k) \approx \frac{m}{\hbar} + \frac{\hbar k^2}{2m}$ when $\alpha_\pm \to 0$. Figure \ref{fig:Plot_w_vs_k_compar} graphically represents the behavior of equation \eqref{eq:Mod_Disp_Relt_Entrop} in comparison with the relativistic frequency dispersion relation \( \omega_{rel} \) and its non-relativistic approximation \( \omega_{cl} \). From this graph, we observe that for small values of the wavenumber (\( k = 2\pi/\lambda \)), all the considered angular frequencies coincide. As \( k \) increases, the modified frequency \( \omega_{-} \) grows faster than the others due to the additional term containing \( k^4 \) and the fact that the deformation parameter \( \alpha_{-} \) is positive. In contrast, the modified frequency \( \omega_{+} \) decreases rapidly as \( k \) increases, since the deformation parameter \( \alpha_{+} \), associated with the non-extensive entropy \( S_{+} \), is negative.

In \cite{XIANG2006519}, it is argued that the emergence of MDR can be attributed to a fluctuating background dominated by quantum effects. This is because the dispersion relation of a particle propagating through a fluctuating spacetime differs from that of a particle moving in a smooth background. This suggests that the non-extensive nature of quantum gravity discussed earlier could be a consequence of spacetime fluctuations at Planck-scale lengths. Thus, it is reasonable to assume that in the classical regime, where spacetime is smooth, the effects of non-extensivity disappear. Unlike the undeformed case, the dispersion relation in \eqref{eq:Mod_Disp_Relt_Entrop} consists of three parts: the first term, \( m/\hbar \), is related to the De Broglie frequency of the particle's rest mass; the second term, \( \hbar k^2 / 2m \), arises from the kinetic energy; and the third term, \( \hbar^3 k^4 / 4m^2 \), can be interpreted as a correction induced by quantum spacetime fluctuations \cite{WHEELER1957604}.

In models such as Loop Quantum Gravity (LQG), Quantum Foam, Hořava-Lifshitz Gravity, and Doubly Special Relativity (DSR), where spacetime is not a smooth classical entity but exhibits fluctuations at Planck scales, particle trajectories are no longer perfectly well-defined \cite{PhysRevD.79.084008,doi:10.1142/S0217751X95000085,doi:10.1142/S0218271802001330,PhysRevD.85.024041,gonzalezmestres1995cosmologicalimplicationspossibleclass}. As well in the quantum mechanical dispersions studied with Matrix theory one also has corrections including powers of $k^4$ \cite{banks97,beckers1} and higher; also a probe graviton dispersion includes these powers \cite{paban98}. A \( k^4 \) term in the dispersion relation introduces a stronger dispersion at high energies, causing particles with different values of \( k \) (momenta) to propagate at different speeds. The group velocity associated with dispersion is given by $v_{g_\pm} \approx \frac{\hbar k}{m} + \alpha_{\pm} \frac{\hbar^3 k^3}{m^2}$. For \( \alpha_{+} < 0 \), the group velocity decreases with \( k \), suggesting a slower propagation of high-energy particles. In contrast, if \( \alpha_{-} > 0 \), the group velocity increases more rapidly with \( k \), causing high-energy particles to travel faster than expected.


Starting from \eqref{Eqq.Eff.Energy.S_{+-}}, similar to what was done in \eqref{Modif.Conmut.Relat.Diment.x.p_pm}, we can calculate the commutation relation between the operator $\hat{T}$ and the effective energy $\hat{E}_\pm$, obtained from entropic principles. This yields
\begin{equation}
    [\hat{T},\hat{E}_\pm]=i\hbar\left(1+\alpha_\pm \hat{E}^2+\cdots\right),\label{eq:Conmut_rel_T_E_expand}
\end{equation}
which is a generalization of \eqref{eq.Usuall_Commut_rela_T_E}, taking into account the quantum gravity effects associated with the non-extensivity of the entropies considered in \eqref{Eqq.Entropy.S_+} and \eqref{Eqq.Entropy.S_-}. Unlike the standard case, tempus and energy aren't longer canonical conjugates. Replacing the deformed Heisenberg algebra \eqref{eq:Conmut_rel_T_E_expand} in the general uncertainty relation \eqref{eq:General_Heisenbrg_ineqlty}, we obtain
\begin{equation}
    \Delta T\Delta E_\pm\geq\frac{\hbar}{2}(1+\alpha_\pm (\Delta E)^2+\cdots).\label{enq:Delt_T_and_Delt_E_mod}
\end{equation}
Similarly to the usual case, we can find the relationship between $\Delta T$ and $\Delta t$ by considering the fact that the operator $\hat{T}$ has not been modified and satisfies the usual algebra with $\hat{E}$. Therefore, for a conservative system, from \eqref{eq:Delta_t_Delta_T_velocty} we have that $\Delta T = \Delta t$. So, \eqref{enq:Delt_T_and_Delt_E_mod} become in
\begin{equation}
    \Delta t\Delta E_\pm\geq\frac{\hbar}{2}(1+\alpha_\pm (\Delta E)^2+\cdots).\label{enq:Uncert_rel_t_E_expand}
\end{equation}
This is the modified uncertainty relation for time and energy, derived from non-extensive entropies. Equation \eqref{enq:Uncert_rel_t_E_expand} is a generalization of \eqref{enq.time_energy_uncertainty_relation} that incorporates high-energy corrections encoded in the non-extensive deformation parameter $\alpha_\pm$. Additionally, if we consider the de Broglie relation for matter waves, then in \eqref{enq:Uncert_rel_t_E_expand} we obtain the modified time-frequency uncertainty relation
\begin{equation}
    \Delta t\Delta \omega_\pm\geq\frac{1}{2}\left(1+\alpha_\pm \hbar^2 (\Delta \omega)^2+\cdots\right),\label{eq:Time_frequ_uncertai_GUP}
\end{equation}
for which we have considered $\Delta E_\pm=\hbar\Delta\omega_\pm$. For example, in \cite{Dodonov_2015}, a Gaussian signal is considered, resulting in $\Delta t=1/\sqrt{2}$ and $\Delta \omega=\sqrt{(\pi-2)/2\pi}$, and therefore the usual uncertainty relation gives $\Delta t\Delta\omega=0.301$. In this case, from \eqref{eq:Time_frequ_uncertai_GUP}, we have $\Delta t\Delta\omega_{+}=0.301-2.38\times 10^{-54}$ and $\Delta t\Delta\omega_{-}=0.301+1.53\times 10^{-54}$. Clearly, in both cases, it can be seen that the first correction term is very small compared to $0.301$. This leads us to interpret that the corrections introduced by the non-extensive entropies $S_{+}$ and $S_{-}$ only play an important role at Planck scales (frequencies), which corroborates the results obtained in \cite{BIZET2023137636}.

Considering the usually relation $\Delta t \Delta E \geq \hbar/2$, it can be concluded from \eqref{enq:Uncert_rel_t_E_expand} that
\begin{equation}
    \Delta E_\pm=\Delta E(1+\alpha_\pm (\Delta E)^2+\cdots).\label{eq:Energy_fluctuation}
\end{equation}
This result can be interpreted in two ways. If the positive deformation parameter \(\alpha_{-}\) is considered, the energy fluctuation increases in the region where the contribution of gravity is significant, that is, at Planck scales, which implies the appearance of an unobservable region $\Delta q$ that cannot be probed \cite{SCARDIGLI199939}. On the other hand, if the negative parameter \(\alpha_{+}\) is considered, the energy fluctuation decreases. Furthermore, as $\alpha_{+} (\Delta E)^2<1$ must be satisfied in \eqref{eq:Energy_fluctuation}, the energy fluctuation reaches its maximum in this scenario. This behavior has been observed in models that consider a crystal-like universe, whose lattice spacing is of the order of the Planck length \cite{PhysRevD.81.084030}.

Now, we intend to calculate the quantum-corrected radiation temperature of a black hole. From \eqref{enq:Uncert_rel_t_E_expand}, we have
\begin{equation}
    \Delta E_\pm\thickapprox\frac{\hbar}{2\Delta t}+\frac{\alpha_\pm\hbar}{2\Delta t}(\Delta E)^2+\cdots.\label{eq:Mod_DeltaE}
\end{equation}
In \cite{scardigli1995some}, a heuristic derivation of the radiation temperature of a black hole is presented by considering particles as real gas at the horizon. Since these particles are moving in a space slice of thickness $\Delta x$, each of them has an uncertainty in the kinetic energy equal to $\Delta E=3k_\beta \Theta/2$, where $\Theta$ is the temperature of this gas of real particles. This provides an estimate of the temperature. We will follow a similar procedure to calculate the contributions of quantum gravity effects to Hawking radiation. For a black hole, we have $\Delta t=2 R^2/GM$ and $\Delta E=\hbar/16GM$, where $R=2GM$ is the Schwarzschild radius, $G$ is the gravitational constant, and $M$ is the mass of the black hole. Thus, in \eqref{eq:Mod_DeltaE}, we have
\begin{equation}
    \Delta E_\pm\thickapprox\frac{\hbar}{16 GM}+\alpha_\pm\left(\frac{\hbar}{16 GM}\right)^3,
\end{equation}
and the corresponding modified Hawking radiation temperature associated with the different statistics imposed by the entropies $S_{+}$ and $S_{-}$ are
\begin{align}
    \Theta_{+}=&\frac{\hbar}{24k_\beta GM}\left[1-\frac{9|\alpha_{+}|k_\beta^2}{4}\left(\frac{\hbar}{24 k_\beta G M}\right)^2\right],\label{eq:Mod_tempe_s+}\\
    \Theta_{-}=&\frac{\hbar}{24k_\beta GM}\left[1+\frac{9\alpha_{-}k_\beta^2}{4}\left(\frac{\hbar}{24 k_\beta G M}\right)^2\right],\label{eq:Mod_tempe_s-}
\end{align}
where $k_\beta$ is the Boltzmann constant. The first term in \eqref{eq:Mod_tempe_s+} and \eqref{eq:Mod_tempe_s-} is the usual Hawking radiation temperature $\Theta=\hbar/24k_\beta GM$ \cite{scardigli1995some}, and the second term contains quantum gravity effects. In the classical limit, where the mass $M$ of the black hole is sufficiently large, the correction terms in \eqref{eq:Mod_tempe_s+} and \eqref{eq:Mod_tempe_s-} are too small to be considered compared to the first term. For example, suppose $M=M_{\text{sun}}=2\times 10^{30} \text{kg}$; the contribution of quantum gravity effects is on the order of $10^{-96}$ compared to $1$ for the first term. On the other hand, if we consider a micro black hole with a mass on the order of the Planck mass $M\sim M_{Pl}$, quantum gravity effects will be significant. This coincides with the observation in \cite{BIZET2023137636} that the modifications are relevant in the quantum gravity realm.  

From \eqref{eq:Mod_tempe_s+}, it can be seen that the radiation temperature for micro black holes, denoted as $\Theta_{+}$, is lower than the Hawking radiation temperature $\Theta$. This result implies that quantum gravity effects related to the non-extensivity associated with entropy $S_{+}$ decrease the black hole's radiation temperature. Conversely, quantum effects due to $S_{-}$ increase the temperature of the quantum micro black hole, as shown in \eqref{eq:Mod_tempe_s-}, compared to the usual Hawking temperature. Several authors by other means have also obtained corrected black hole temperatures in different contexts \cite{obregon2001entropy, BARGUENO201515, Vagenas_2017,PhysRevD.104.066012, farmany2011generalized, MYUNG2007393, NOUICER200763}.\par


We can extend the commutation relation \eqref{eq:Conmut_rel_T_E_expand} to a form analogous to \eqref{Modif.Conmut.Relat.Diment.x.p_pm.Fam} considering all corrections terms in $\alpha_\pm$, or equivalently, by setting $\alpha_\pm \hat{E}_\pm^2=\alpha_\pm \hat{E}^2+\cdots$. Thus, we have
\begin{equation}
    [\hat{T},\hat{E}_\pm]=i\hbar\left(1+\alpha_\pm \hat{E}_\pm^2\right),\label{eq:Tempus_Energy_Conm_rel_Mod}
\end{equation}
and the corresponding uncertainty relation
\begin{eqnarray}
    \Delta T\Delta E_\pm\geq\frac{\hbar}{2}\left(1+\alpha_\pm\left(\Delta E_\pm\right)^2\right).\label{eq:Gener_Uncer_relat_T_E}
\end{eqnarray}
Using the same argument as in \eqref{enq:Uncert_rel_t_E_expand}, we can express the modified uncertainty relation between time and energy from \eqref{eq:Gener_Uncer_relat_T_E}, which is as follows
\begin{equation}
    \Delta t\Delta E_\pm\geq\frac{\hbar}{2}\left(1+\alpha_\pm\left(\Delta E_\pm\right)^2\right),\label{eq:Gener_Uncer_relat_t_E}
\end{equation}
that is the generalized version of \eqref{enq.time_energy_uncertainty_relation}, the energy-time Generalized Uncertainty Relation. As mentioned earlier, the deformation parameter $\alpha_\pm$ has two possible values, one positive, $\alpha_-$, and the other negative, $\alpha_+$, depending on the statistics considered. Choosing the positive parameter in \eqref{eq:Gener_Uncer_relat_t_E}, associated with the entropy $S_{-}$, leads to a minimum time interval
\begin{equation}
    (\Delta t)_{\text{min}}=t_{\text{min}}=\hbar\sqrt{\alpha_{-}}.\label{eq:Minimal_time}
\end{equation}
On the contrary, if one chooses the negative value of the deformation parameter denoted by $\alpha_{+}$, associated with the entropy $S_{+}$,  a maximum energy width is found
\begin{equation}
    (\Delta E_{+})_{\text{max}}=E_{\text{max}}=\frac{1}{\sqrt{|\alpha_{+}|}}.\label{eq:Maximal_energy}
\end{equation}
Taking into account that in the Planck regimen $\alpha_\pm=\alpha_\pm^0/M_{Pl}^2$, then the minimum time in \eqref{eq:Minimal_time} is on the order of the Planck time $\tau_{Pl}$, $t_{\text{min}}=\tau_{Pl}\sqrt{\alpha_-^{0}}$. In \eqref{eq:Maximal_energy}, this implies that the maximum energy is on the order of the Planck energy $M_{Pl}$, i.e., $E_{\text{max}}=M_{Pl}/\sqrt{|\alpha_+^{0}|}$. Notice first that the extreme magnitudes are of the order of the Planck scale. Also, is important to see that from the different statistics we obtain minimum time or maximum energy. This is given by the signs of the $\alpha_{\pm}$, meaning that one class of entropies corresponds to one case and the other entropy to the other one.

 Non-extensive statistics characterized by entropies $S_{\pm}$ lead to distinct physical behaviors. For example, in \eqref{eq:Minimal_time} and \eqref{eq:Mod_tempe_s-}, $S_{-}$ leads to a minimum time and an increase in the radiation temperature of a mini black hole, while in \eqref{eq:Maximal_energy} and \eqref{eq:Mod_tempe_s+}, $S_{+}$ leads to maximum energy and a decrease in the Hawking temperature. This fact was already noted in \cite{BIZET2023137636} because the statistics associated with $S_+$ and $S_-$ have different characteristics. The first entropy gives rise to an effective potential related to an effective repulsive contribution term, while the second one gives rise to an effective attractive contribution \cite{GILVILLEGAS2017364}.

 It is evident that the results we obtained in this part of the work align with those presented in \eqref{enq:Delta_t_Delta_E_GUP} concerning micro-black hole formation, particularly with the results derived from the entropy $S_{-}$. This strengthens our argument that the effects of non-extensivity, encoded in the entropies $S_{\pm}$, are significant at the order of the Planck scale and therefore represent a signature of quantum gravity.

 In \cite{Dodonov_2015}, the decay laws are discussed based on their relationship with the energy spectrum of the system \cite{krylov1947two}. Due to \eqref{eq:Gener_Uncer_relat_t_E}, the representation of the energy eigenstates must change, which would result in a modified non-decay amplitude $\chi(t)$ \cite{fleming1973unitarity}, and consequently, the decay law $Q(t)$. In future work, we will analyze these and other related issues more carefully.

\section{Conclusions}
\label{sec:Conclusions}
With the goal of introducing the quantum effects of gravity into measurement processes, in an effective manner, the uncertainty relation between position and momentum is modified \eqref{enq:Dq_Dp_GUP_Scardigli}. This modification is named the Generalized Uncertainty Principle (GUP). Various approaches have demonstrated the emergence of GUP \cite{G_Veneziano_1986, MAGGIORE199365, SCARDIGLI199939}. Notably, an intriguing connection has been established between non extensive entropies and GUP, as $S_\pm$ in \eqref{Eqq.Entropy.S_+}-\eqref{Eqq.Entropy.S_-} that depending only on the probability, \cite{BIZET2023137636} (see \eqref{enq.GUP_q_p}). This connection allows us to interpret that quantum gravity exhibits non-extensive behavior, which can be quantified through non-extensive entropies. The statistics associated with $S_{+}$ and $S_{-}$ yields distinct interpretations. For example, those related to $S_{+}$ result in a maximum uncertainty for the momentum, while those associated with $S_{-}$ lead to a minimum uncertainty for the position. Furthermore, $S_{+}$ gives rise to an effective potential correlated with a repulsive interaction, whereas $S_{-}$ gives rise to an effective attractive interaction \cite{GILVILLEGAS2017364}.

Spacetime fluctuations in the high-energy quantum regime enable the formation of micro black holes, whose gravitational effects modify Heisenberg's uncertainty relation~\cite{SCARDIGLI199939}. Similarly, when these effects are taken into account, the energy-time uncertainty relation is also modified (see \eqref{enq:Delta_t_Delta_E_GUP}). As a result, the time interval required for the energy of a state to dissipate acquires a minimum value (on the order of the Planck time), after which it increases due to the formation of the micro-black hole. It is important to emphasize that these effects are significant only at scales close to the Planck scale.

Using the probability distributions derived from the generalized entropies $S_\pm$, \eqref{6}, we obtain two effective Hamiltonians, labeled in \eqref{eq:Eff_Hamiltonian}, which include correction terms at the Planck scale. At high-energy and high-velocity regimes, the energy and the momentum exhibit similar characteristics; therefore we express the effective Hamiltonian in its relativistic form \eqref{Eqq.Eff.Energy.S_{+-}}, in which the concepts of relativity and quantum theory are combined. As a result, from the generalized relativistic energy, a Modified Dispersion Relation (MDR) \eqref{eq:MDR_entropy} is derived. Unlike other proposals, such as \cite{XIANG2006519}, this relation is obtained starting from entropic principles. This suggests that the non-extensive nature exhibited by quantum gravity originates from spacetime quantum fluctuations at Planck scales. Thus, it is reasonable to assume that in the classical regime, where spacetime is smooth, the effects of non-extensivity vanish. Finally, we can conclude that non-extensivity is a signature of quantum gravity \cite{BIZET2023137636}.

We employ the definition of the tempus operator \cite{PhysRevA.50.933}, computing the commutator with the effective energy operator \eqref{Eqq.Eff.Energy.S_{+-}}. This yields a modified commutation relations \eqref{eq:Conmut_rel_T_E_expand}, leading to the generalized energy-time uncertainty relation \eqref{enq:Uncert_rel_t_E_expand}. Consistently with previous findings, the corrections arising from $S_{+}$ and $S_{-}$ exhibit differences. The deformation parameter $\alpha_+$, derived from $S_{+}$, results in a decrease in energy fluctuation and the radiation temperature of a micro black hole, see \eqref{eq:Energy_fluctuation} and \eqref{eq:Mod_tempe_s+}, as well as in maximum fluctuation in energy \eqref{eq:Maximal_energy}. Conversely, the deformation parameter $\alpha_{-}$ derived from $S_{-}$ leads to an increase in the energy fluctuation and radiation temperature, as indicated in \eqref{eq:Energy_fluctuation} and \eqref{eq:Mod_tempe_s-}, as well as to the emergence of a minimum time interval \eqref{eq:Minimal_time}, which represents the minimum time required for the energy $\Delta E_{-}$, concentrated in the $\Delta x_{-}$ region, to escape from that region. In the Planck regime, the minimum time interval is in the order of the Planck time, and the maximum fluctuation in energy is in the order of the Planck energy.

We want to emphasize that, just as it was demonstrated that incorporating gravity into measurement processes leads to modifications in the position-momentum \eqref{enq:Dq_Dp_GUP_Scardigli} and energy-time \eqref{enq:Delta_t_Delta_E_GUP} uncertainty relations, non-extensivity naturally arises at energy scales where quantum gravity effects are significant, modifying the uncertainty relations \eqref{enq.GUP_q_p} and \eqref{eq:Gener_Uncer_relat_t_E}. These modifications play a crucial role at energy and length scales near the Planck scale and become negligible in the classical regime.

The outcomes presented in this work strengthen the proposition that non-extensivity in entropy is a manifestation of quantum gravity \cite{BIZET2023137636}. The quantum corrections originating from the generalized entropies $S_\pm$ prove to be substantial on both Planck scales (GUP). Grasping these findings unveils considerable potential for applying non-extensive entropies to understand phenomena entailing quantum gravity, a subject that will be explored further in forthcoming research.

\acknowledgments
N. C. B and W. Y thank the support of the University of Guanajuato grants:  Convocatoria Institucional de Investigación Científica (CIIC) 251/2024 ``Teorías efectivas de cuerdas y exploraciones de aprendizaje de máquina'',  CIIC 264/2022 ``Geometría de dimensiones extras en teoría de cuerdas y sus aplicaciones físicas''. NCB would like to thank the grant  by Consejo Nacional de Humanidades Ciencias y Tecnologías (CONAHCyT) A-1-S-37752,``Teorías efectivas de cuerdas y sus aplicaciones a la física de partículas y cosmología'' and  University of Guanajuato grant CIIC 224/2023 ``Conjeturas de gravedad cuántica y el paisaje de la teoría de cuerdas''. NCB would like to acknowledge support from the ICTP through the Associates Programme (2023-2029), and the Isaac Newton Institute for Mathematical Sciences, Cambridge, for support and hospitality during the programme ``Black holes: bridges between number theory and holographic quantum information'' where work on this paper was undertaken. This work was supported by EPSRC grant no EP/R014604/1.  O. O and W. Y thanks to the grant by the University of Guanajuato CIIC 168/2023 ``Non-extensive Entropies Independent of Parameters'' and CIIC 156/2024 ``Generalized Uncertainty Principle, Non-extensive Entropies, and General Relativity'', as well as the SECIHTI grant CBF2023-2024-2923 ``Implications of the Generalized Uncertainty Principle (GUP) in Quantum Cosmology, Gravitation, and its Connection with Non-extensive Entropies''. W. Y is supported by SECIHTI/Estancias Posdoctorales por México.




\bibliographystyle{JHEP}

\bibliography{biblio}
\end{document}